\documentclass[8.5pt,twoside,twocolumn]{article}
\usepackage[super,sort&compress,comma]{natbib}
\usepackage{mhchem}
\usepackage{times,mathptmx}
\usepackage{sectsty}
\usepackage{balance}

\usepackage{graphicx} 
\usepackage{lastpage}
\usepackage[format=plain,justification=raggedright,singlelinecheck=false,font=small,labelfont=bf,labelsep=space]{caption}
\usepackage{fancyhdr}
\begin{document}

\thispagestyle{plain}
\fancypagestyle{plain}{
\fancyhead[L]{\includegraphics[height=8pt]{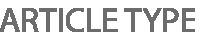}}
\fancyhead[C]{\hspace{-1cm}\includegraphics[height=20pt]{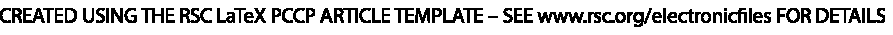}}
\fancyhead[R]{\includegraphics[height=10pt]{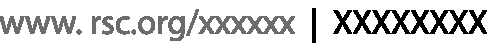}\vspace{-0.2cm}}
\renewcommand{\headrulewidth}{1pt}}
\renewcommand{\thefootnote}{\fnsymbol{footnote}}
\renewcommand\footnoterule{\vspace*{1pt}%
\hrule width 3.4in height 0.4pt \vspace*{5pt}}
\setcounter{secnumdepth}{5}

\makeatletter
\def\subsubsection{\@startsection{subsubsection}{3}{10pt}{-1.25ex plus -1ex minus -.1ex}{0ex plus 0ex}{\normalsize\bf}}
\def\paragraph{\@startsection{paragraph}{4}{10pt}{-1.25ex plus -1ex minus -.1ex}{0ex plus 0ex}{\normalsize\textit}}
\renewcommand\@biblabel[1]{#1}
\renewcommand\@makefntext[1]%
{\noindent\makebox[0pt][r]{\@thefnmark\,}#1}
\makeatother
\renewcommand{\figurename}{\small{Fig.}~}
\sectionfont{\large}
\subsectionfont{\normalsize}

\fancyfoot{}
\fancyfoot[LO,RE]{\vspace{-7pt}\includegraphics[height=9pt]{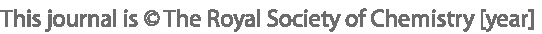}}
\fancyfoot[CO]{\vspace{-7.2pt}\hspace{12.2cm}\includegraphics{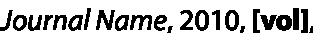}}
\fancyfoot[CE]{\vspace{-7.5pt}\hspace{-13.5cm}\includegraphics{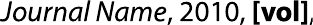}}
\fancyfoot[RO]{\footnotesize{\sffamily{1--\pageref{LastPage} ~\textbar  \hspace{2pt}\thepage}}}
\fancyfoot[LE]{\footnotesize{\sffamily{\thepage~\textbar\hspace{3.45cm} 1--\pageref{LastPage}}}}
\fancyhead{}
\renewcommand{\headrulewidth}{1pt}
\renewcommand{\footrulewidth}{1pt}
\setlength{\arrayrulewidth}{1pt}
\setlength{\columnsep}{6.5mm}
\setlength\bibsep{1pt}

\twocolumn[
  \begin{@twocolumnfalse}
\noindent\LARGE{\textbf{Magnetization, crystal structure and anisotropic thermal expansion of single-crystal SrEr$_2$O$_4$}}
\vspace{0.6cm}

\noindent\large{\textbf{
Hai-Feng Li,$^{\ast}$\textit{$^{a,b}$}
Andrew Wildes,\textit{$^{c}$}
Binyang Hou,\textit{$^{d}$}
Cong Zhang,\textit{$^{b}$}
Berthold Schmitz,\textit{$^{e}$}
Paul Meuffels,\textit{$^{f}$}
Georg Roth\textit{$^{b}$} and
Thomas Br$\ddot{\texttt{u}}$ckel,\textit{$^{e}$}
}}\vspace{0.5cm}

\noindent\textit{\small{\textbf{Received Xth XXXXXXXXXX 2014, Accepted Xth XXXXXXXXX 2014\newline
First published on the web Xth XXXXXXXXXX 200X}}}

\noindent \textbf{\small{DOI: 10.1039/b000000x}}
\vspace{0.6cm}

\noindent
\normalsize{
The magnetization, crystal structure, and thermal expansion of a nearly stoichiometric Sr$_{1.04(3)}$Er$_{2.09(6)}$O$_{4.00(1)}$ single crystal have been studied by PPMS measurements and in-house and high-resolution synchrotron X-ray powder diffraction. No evidence was detected for any structural phase transitions even up to 500 K. The average thermal expansions of lattice constants and unit-cell volume are consistent with the first-order Gr$\ddot{\texttt{u}}$neisen approximations taking into account only the phonon contributions for an insulator, displaying an anisotropic character along the crystallographic \emph{a}, \emph{b}, and \emph{c} axes. Our magnetization measurements indicate that obvious magnetic frustration appears below $\sim$15 K, and antiferromagnetic correlations may persist up to 300 K.
}
\vspace{0.5cm}
 \end{@twocolumnfalse}
  ]

\footnotetext{\textit{$^{a}$~J$\ddot{u}$lich Centre for Neutron Science JCNS, Forschungszentrum J$\ddot{u}$lich GmbH, Outstation at Institut Laue-Langevin, Bo$\hat{\imath}$te Postale 156, F-38042 Grenoble Cedex 9, France. Fax: +49 2461 612610; Tel: +49 2461 614750; E-mail: h.li@fz-juelich.de}}
\footnotetext{\textit{$^{b}$~Institut f$\ddot{u}$r Kristallographie der RWTH Aachen University, D-52056 Aachen, Germany}}
\footnotetext{\textit{$^{c}$~Institut Laue-Langevin, Bo$\hat{\imath}$te Postale 156, F-38042 Grenoble Cedex 9, France}}
\footnotetext{\textit{$^{d}$~European Synchrotron Radiation Facility, Bo$\hat{\imath}$te Postale 220, F-38043 Grenoble Cedex, France}}
\footnotetext{\textit{$^{e}$~J$\ddot{u}$lich Centre for Neutron Science JCNS and Peter Gr$\ddot{u}$nberg Institut PGI, JARA-FIT, Forschungszentrum J$\ddot{u}$lich GmbH, D-52425 J$\ddot{u}$lich, Germany}}
\footnotetext{\textit{$^{f}$~Peter Gr$\ddot{u}$nberg Institut PGI and JARA-FIT, Forschungszentrum J$\ddot{u}$lich GmbH, D-52425 J$\ddot{u}$lich, Germany}}

\section{Introduction}

The existence of competing Hamiltonian terms, \emph{e.g.}, between single-ion anisotropy and spin-spin interactions, or of competing spin-spin interactions, \emph{e.g.}, between next-nearest spin neighbours, often leads to a large ground-state degeneracy because these competing energy components sometimes cannot minimize simultaneously \cite{Diep2004, Lacroix2011}. In this instance, novel ground states such as spin liquid, spin ice, cooperative paramagnetism, or magnetic Coulomb phase based on magnetic monopole excitations may emerge in frustrated magnets, providing an excellent testing ground for approximations and theories \cite{Diep2004, Lacroix2011, Harris1997, Bramwell2001, Ramirez2003, Moessner2006, Han2008, Morris2009, Gingras2009, Toth2012}.

\begin{figure}[h]
\centering \includegraphics[width = 0.44\textwidth] {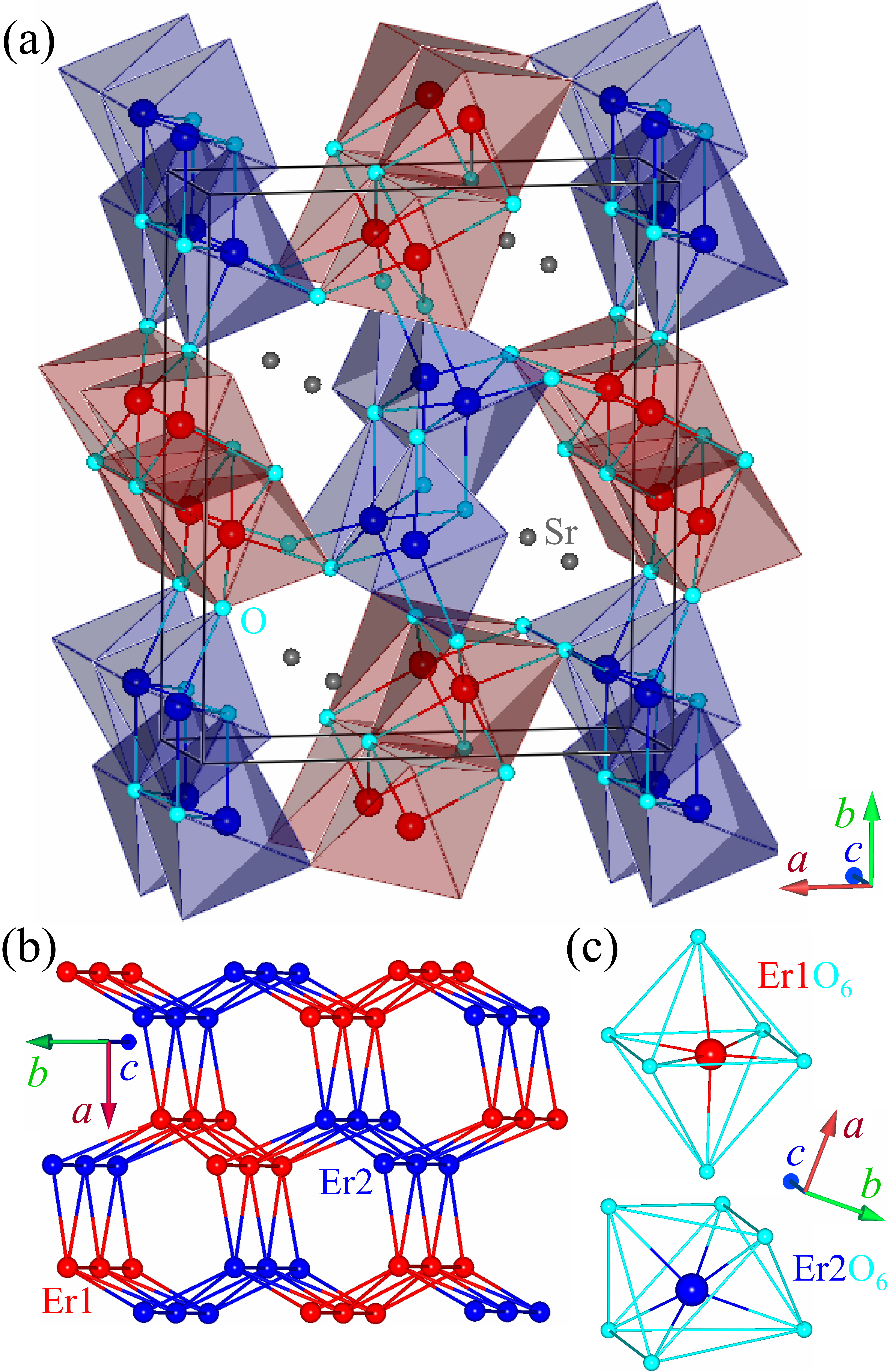}
\caption{
(a) Crystal structure (space group $Pnam$) of SrEr$_2$O$_4$ in one unit cell (solid lines) as refined from the I11 (Diamond) data measured at 80 K.
(b) Bent honeycombs that are composed by the Er$_6$ (3Er1 plus 3Er2) hexagons are stacked along the crystallographic $c$ axis, with the shortest Er-Er bonds along that direction.
(c) The corresponding octahedra of Er1O$_6$ and Er2O$_6$.
All oxygen sites (O1, O2, O3, and O4 as listed in Table 2) are schematically illustrated with the same color code.
}
\label{Fig1}
\end{figure}

The compound SrEr$_2$O$_4$ is one of the geometrically-frustrated magnets in the family of Sr$RE_2$O$_4$ ($RE =$ Gd, Tb, Dy, Ho, Er, Tm, and Yb) compounds \cite{Barry1967, Paletta1968, Chaker2003, Karunadasa2005, Doi2006, Jin2008, Hirose2009, Lia2009, Ghosh2011, Ofer2011, Quintero2012, Hayes2011, Li2014Tm, Fennell2014, Li2014Tb, Besara2014, Aczel2014}. They crystallise with an unusual orthorhombic (space group $Pnam$, $Z =$ 4) structure \cite{Pepin1981} (Fig.~\ref{Fig1}(a)), in which two inequivalent crystallographic sites accommodate the $RE^{3+}$ ions. Therefore, there are two types of Er$^{3+}$ ions (Er1 and Er2), as shown in Fig.~\ref{Fig1}(b), residing in different crystallographic environments and forming two different octahedra (Er1O$_6$ and Er2O$_6$) as shown in Fig.~\ref{Fig1}(c). This may strongly influence their respective magnetic states by virtue of different crystal field effects \cite{Fennell2014}. The neighbour Er1O$_6$ or Er2O$_6$ octahedra share edges. The Er1O$_6$ octahedron connects with its neighbour Er2O$_6$ octahedra by way of sharing their in-plane and spatial corners, thereby forming a 3D network of the ErO$_6$ octahedra. The shortest Er-Er bonds form Er chains along the crystallographic \emph{c} axis (Fig.~\ref{Fig1}(b)), indicating that the strongest magnetic coupling is probably along that direction \cite{Li2014Tb}. Three Er1 and three Er2 ions comprise the bent Er$_6$ honeycombs normal to the crystallographic \emph{c} axis (Fig.~\ref{Fig1}(b)). With this crystallographic arrangement, the low coordinate number of Er$^{3+}$ ions and the competing magnetic couplings between Er$^{3+}$ chains may lead to a geometrical frustration.

It was reported that there exist two types of magnetic order in the compound SrEr$_2$O$_4$ \cite{Hayes2011}. One is a long-range magnetic order with moment direction along the crystallographic \emph{c} axis and antiferromagnetic (AFM) phase transition temperature at $T_\texttt{N} =$ 0.75 K. Another is a short-range magnetic order, which persists up to much higher temperatures (than $T_\texttt{N}$) with moment direction parallel to the crystallographic \emph{a} axis. So far, the crystallographic origins of the two kinds of magnetic order are hard to be distinguished. They may be from either Er1 or Er2, or both sites.

Determining magnetic coupling mechanisms always represents a critical step toward a complete understanding of the relevant magnetic frustrations, for which one prerequisite is to accurately analyze the related crystal structure. As suspected, there may exist possible structural phase transitions as function of temperature in the compound SrEr$_2$O$_4$ \cite{Li2014Tb, Li2014Tm}. To our knowledge, no structural study of SrEr$_2$O$_4$ has been performed by high-resolution synchrotron X-ray diffraction. In addition, all reported structural parameters were limited to a certain temperature point below 300 K. Therefore, the detailed thermal expansion of the compound SrEr$_2$O$_4$ with temperature has not been determined yet.

In this paper, we report magnetic characterizations and powder in-house and synchrotron X-ray diffraction studies of a Sr$_{1.04(3)}$Er$_{2.09(6)}$O$_{4.00(1)}$ single crystal. The space group symmetry of the crystal structure is invariant with temperature up to 500 K. There exists an anisotropic change in lattice constants along the crystallographic \emph{a}, \emph{b}, and \emph{c} axes, with the largest relative thermal expansion being along the \emph{a} axis. The Curie-Weiss (CW) temperature is negative, \emph{e.g.}, $\sim$-15.94 K in case of the zero-field cooling (ZFC), indicating a strong net AFM coupling strength. Appreciable magnetic frustration is perceived below $\sim$15 K so that the measured spin-moment size at 2 K and 9 T, \emph{i.e.}, 4.3(5) $\mu_\texttt{B}$, may be from only one of the two inequivalent Er$^{3+}$ crystallographic sites.

\section{Experimental}

The sample synthesis is similar to that reported previously \cite{Li2014Tb, Li2008}. We quantitatively determine chemical compositions of the studied single crystal by inductively coupled plasma with optical emission spectroscopy (ICP-OES) analysis. We measured the ZFC and field cooling (FC) dc magnetization as a function of temperature from 2 to 300 K at 500 Oe, and \emph{versus} applied magnetic field (up to 9 T) at 2-50 K (detailed temperature points are listed in Fig.~\ref{Fig3}), using a commercial physical property measurement system.

The in-house X-ray powder-diffraction (XRPD) was carried out on a diffractometer, employing the copper $K_{\alpha 1}$ = 1.5406(9) {\AA} radiation, with a 2$\theta$ step size of 0.005$^\circ$ in a transmission geometry from 15 to 300 K. The bulk sample was gently ground into powder and then pressed onto a thin Mylar film to form a flat surface.

High-resolution synchrotron X-ray powder-diffraction (SXRPD) patterns for structure solution were collected over the 2$\theta$ range 0-150$^{\circ}$ at 80, 298, and 500 K using beamline I11 \cite{Parker2011} at Diamond Light Source, Didcot, UK. The calibrated X-ray wavelength was $\lambda$ = 0.82703(6) {\AA} with a detector zero angle offset of 0.00496(1)$^{\circ}$, as determined by our Rietveld refinements of the data. A powdered sample of SrEr$_2$O$_4$ was loaded onto the outside surface of a 0.3 mm diameter borosilicate glass capillary tube by attaching a thin layer of hand cream. The tube was rolling to minimise the effects of absorption and preferred orientation during data collection from an even coat of sample. The beamline comprises a transmission geometry X-ray instrument with a wide range position sensitive detector.

All powder diffraction data were analyzed by the Fullprof Suite \cite{fullprof}. The peak profile shape was modeled with a Pseudo-Voigt function. We refined the background with a linear interpolation between automatically-selected background points. The wavelength, scale factor, zero shift, peak shape parameters, asymmetry, lattice parameters, atomic positions, isotropic thermal parameter \emph{B}, as well as the preferred orientation, \emph{etc}., were all refined.

The samples used for magnetization measurements and powder-diffraction studies are pulverized single-crystalline SrEr$_2$O$_4$ from the same ingot synthesized in a single growth.

\begin{figure}[h]
\centering \includegraphics[width = 0.44\textwidth] {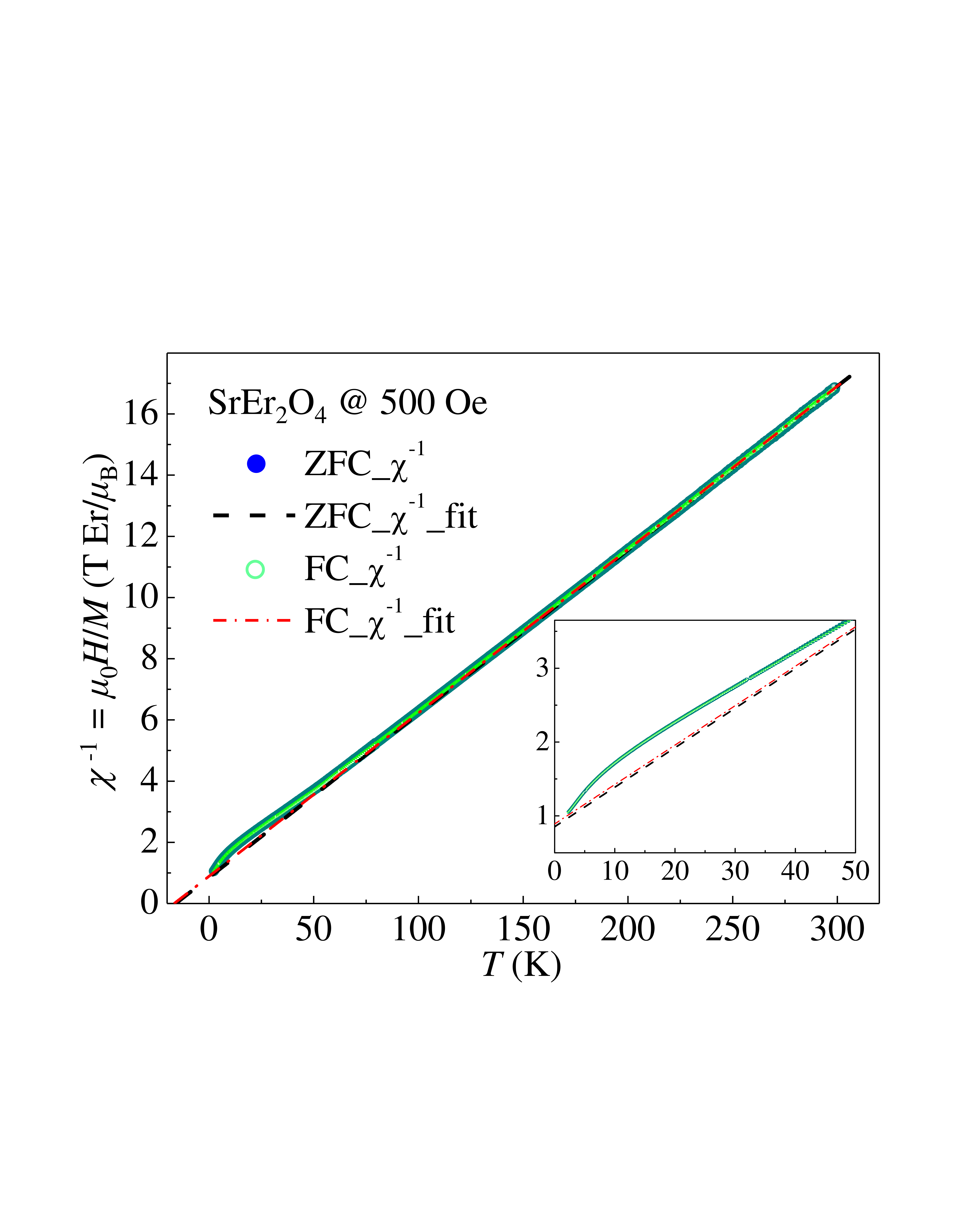}
\caption{
Inverse magnetic susceptibility $\chi^{-1}$ (solid and void circles) deduced from the ZFC and FC magnetization ($M$, normalized to a single Er$^{3+}$ ion) measurements as a function of temperature from 2 to 300 K at 500 Oe. Due to the high density of data points ($\sim$16000), they almost overlap in the entire temperature range. The dashed and dash-dotted lines are fits to the data (200-300 K, $\sim$5600 points) with a CW law as described in the text. The fit results are listed in Table 1. Inset enlarges the most interesting part in the temperature range of 0-50 K.
}
\label{Fig2}
\end{figure}

\begin{figure}[h]
\centering \includegraphics[width = 0.44\textwidth] {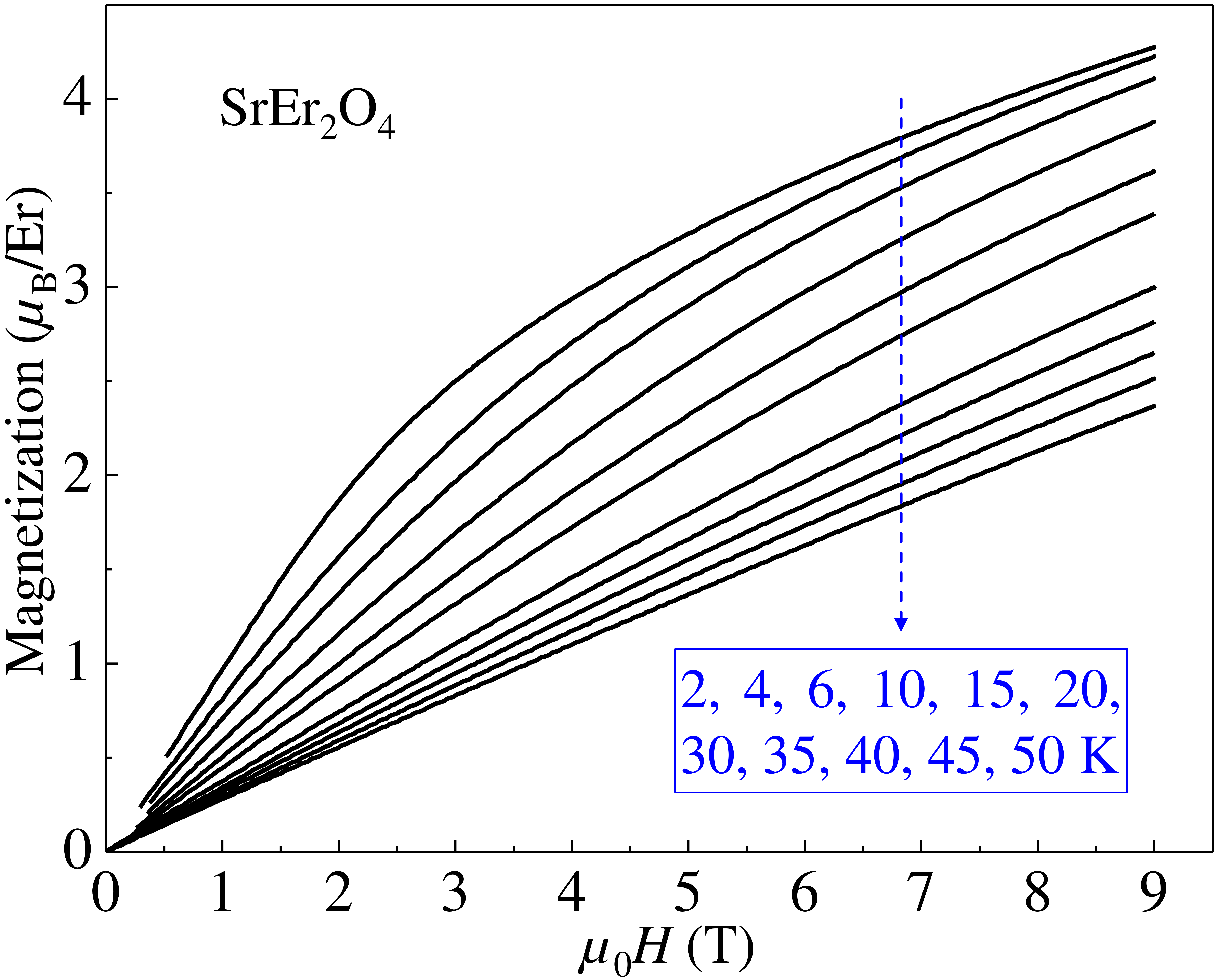}
\caption{
FC magnetization that is normalized to a single Er$^{3+}$ ion \emph{versus} applied magnetic field up to 9 T at temperature points as marked.
}
\label{Fig3}
\end{figure}

\section{Results and discussion}
\subsection{ICP-OES measurements}

The chemical compositions of the studied single crystal were determined as Sr$_{1.04(3)}$Er$_{2.09(6)}$O$_{4.00(1)}$ by our ICP-OES measurements, which indicates that the grown crystal is almost stoichiometric within the experimental accuracy.

It is pointed out that while normalizing the measured magnetization to a single Er$^{3+}$ ion and during the process of refining the collected XRPD and SXRPD patterns, we used the measured stoichiometry by ICP-OES for the normalization and site occupancies in the refinements, respectively.

\subsection{Magnetization \emph{versus} temperature}

We measured the ZFC and FC magnetization ($M$) of a small piece (4.540(1) mg) of the single crystal at 500 Oe. The extracted inverse magnetic susceptibility, \emph{i.e.}, $\chi^{-1} = \mu_0H/M$, is displayed in Fig.~\ref{Fig2}, which theoretically observes the CW law in a paramagnetic (PM) state, \emph{i.e.},
\begin{eqnarray}
\chi(T) = \frac{C}{T - \theta_{\rm CW}} = \frac{N_\texttt{A} M^2_{\rm eff}}{3k_\texttt{B}(T - \theta_{\rm CW})},
\label{CW}
\end{eqnarray}
where \emph{C} is the Curie constant, $\theta_{\rm CW}$ is the CW temperature, $M_{\rm eff}$ is the effective PM moment, $N_\texttt{A}$ = 6.022 $\times$ 10$^{23}$ mol$^{-1}$ is the Avogadro's number, and $k_\texttt{B}$ = 1.38062 $\times$ 10$^{-23}$ J K$^{-1}$ is the Boltzmann constant. We fit the high-temperature (200-300 K) data points with eqn~(\ref{CW}) and extrapolated the fits down to low temperatures (from -16.69 to 200 K), shown as the dashed (ZFC) and dash-dotted (FC) lines in Fig.~\ref{Fig2}. The deduced effective PM moments, $M^{\rm eff}_{\rm mea}$, and CW temperatures are listed in Table 1.

\begin{table}[h]
\small
\caption{\ Quantum numbers of Er$^{3+}$ ions in SrEr$_2$O$_4$: spin \emph{S}, orbital \emph{L}, total angular momentum \emph{J}, and Land$\acute{\texttt{e}}$ factor $g_J$ as well as the ground-state term $^{2S+1}L_J$. Theoretical (theo) and measured (mea) values of the ZFC and FC effective (eff) and FC saturation (sat) Er$^{3+}$ moments, and the corresponding CW temperatures ($\theta_{\texttt{CW}}$) are all listed. Number in parenthesis is the estimated standard deviation of the last significant digit.}
\label{tab1}
\begin{tabular*}{0.5\textwidth}{@{\extracolsep{\fill}}lcr}
\hline
\multicolumn{3}{c} {Single-crystal SrEr$_2$O$_4$}                                                                               \\*
\hline
4$f$ ion                                                                                    &          &    Er$^{3+}$           \\*
4$f^\texttt{n}$                                                                             &          &    11                  \\*
$S$                                                                                         &          &    3/2                 \\*
$L$                                                                                         &          &    6                   \\*
$J = L + S$ (Hund{\textquoteright}s rule for free Er$^{3+}$)                                &          &    15/2                \\*
$g_J$                                                                                       &          &    1.2                 \\*
$^{2S+1}L_J$                                                                                &          &    $^4I_{15/2}$        \\*
$M^{\texttt{eff}}_{\texttt{theo}} = g_J \sqrt{J(J+1)}$      $(\mu_\texttt{B})$              &          &    $\sim$9.58          \\*
$M^{\texttt{sat}}_{\texttt{theo}} = g_J J$ $(\mu_\texttt{B})$                               &          &    9.0                 \\*
\hline
$M^{\texttt{ZFC-eff}}_{\texttt{mea}}/\texttt{Er}^{3+}$ $(\mu_\texttt{B})$                   &          &    9.13(1)             \\*
$M^{\texttt{FC-eff}}_{\texttt{mea}}/\texttt{Er}^{3+}$ $(\mu_\texttt{B})$                    &          &    9.14(1)             \\*
$M^{\texttt{FC}}_{\texttt{mea}}/\texttt{Er}^{3+}$ (2 K, 9 T) $(\mu_\texttt{B})$             &          &    4.3(5)              \\*
$\theta^{\texttt{ZFC}}_{\texttt{CW}}$ (K)                                                   &          &    -15.94(3)           \\*
$\theta^{\texttt{FC}}_{\texttt{CW}}$ (K)                                                    &          &    -16.69(3)           \\*
\hline
\end{tabular*}
\end{table}

Upon cooling, at $\sim$150 K, the measured $\chi^{-1}$ turns clearly upward from the CW estimations, indicating a possible onset of distinguishable AFM correlations. This deviation becomes larger and larger as temperature decreases until $\sim$15 K, accompanied in principle by a progressive increase in the AFM correlations. Below 15 K, as temperature decreases, the difference between measured and CW-estimated $\chi^{-1}$ gets smaller and smaller. At 2 K, the lowest temperature point for the magnetization measurements, they even coincide with each other. This observation indicates that obvious magnetic frustration forms below $\sim$15 K. The lower the temperature is, the more visible the effect of spin frustrations becomes. This is consistent with our field-dependent magnetization measurements (Fig.~\ref{Fig3}) as discussed below.

\begin{figure}[h]
\centering \includegraphics[width = 0.44\textwidth] {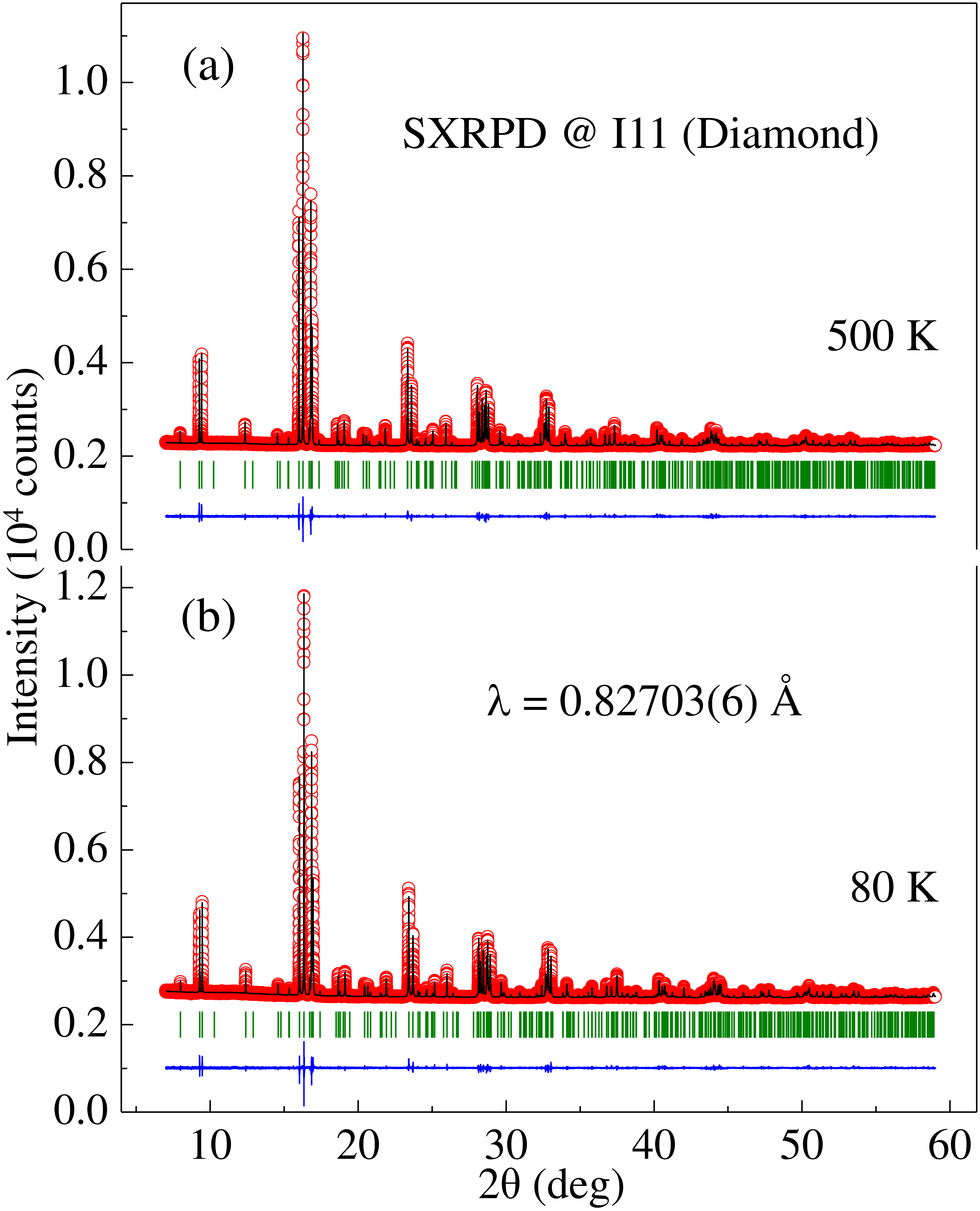}
\caption{
Observed (circles) and calculated (solid lines) SXRPD patterns from the study using I11 (Diamond) at 500 K (a) and 80 K (b). The vertical bars in each panel mark the positions of nuclear Bragg reflections of SrEr$_2$O$_4$. The lower curves represent the difference between observed and calculated patterns.
}
\label{Fig4}
\end{figure}

\begin{table}[h]
\small
\caption{\ Refined structural parameters (lattice constants \emph{a}, \emph{b}, and \emph{c}, unit-cell volume \emph{V}, atomic positions, Debye-Waller thermal parameter \emph{B}), and the corresponding goodness of refinements of the SXRPD data measured at 80, 298, and 500 K using I11 (Diamond). Number in parenthesis is the estimated standard deviation of the last significant digit.}
\label{tab2}
\begin{tabular*}{0.5\textwidth}{@{\extracolsep{\fill}}lccc}
\hline
   \multicolumn{4}{c} {SrEr$_2$O$_4$ (Orthorhombic, \emph{Pnam}, $Z = 4$)}         \\
\hline
\emph{T} (K)           &        80         &       298         &      500          \\
\hline
$a$ ({\AA})            &    10.01660(2)    &    10.03860(2)    &    10.06411(2)    \\
$b$ ({\AA})            &    11.85413(2)    &    11.86714(2)    &    11.88430(2)    \\
$c$ ({\AA})            &    3.38484(1)     &    3.38952(1)     &    3.39571(1)     \\
$V$ ({\AA}$^3$)        &    401.910(1)     &    403.791(1)     &    406.144(1)     \\
\hline
              \multicolumn{4}{c} {Wyckoff site 4c: (\emph{x}, \emph{y}, 0.25)}     \\
$x$ (Sr)               &    0.7527(1)      &    0.7525(1)      &    0.7527(1)      \\
$y$ (Sr)               &    0.6500(1)      &    0.6498(1)      &    0.6496(1)      \\
$x$ (Er1)              &    0.4227(1)      &    0.4228(1)      &    0.4227(1)      \\
$y$ (Er1)              &    0.1101(1)      &    0.1103(1)      &    0.1105(1)      \\
$x$ (Er2)              &    0.4232(1)      &    0.4233(1)      &    0.4233(1)      \\
$y$ (Er2)              &    0.6118(1)      &    0.6118(1)      &    0.6119(1)      \\
$x$ (O1)               &    0.2157(6)      &    0.2126(5)      &    0.2130(5)      \\
$y$ (O1)               &    0.1755(4)      &    0.1737(4)      &    0.1737(4)      \\
$x$ (O2)               &    0.1244(5)      &    0.1236(5)      &    0.1240(5)      \\
$y$ (O2)               &    0.4821(5)      &    0.4837(4)      &    0.4853(4)      \\
$x$ (O3)               &    0.5104(6)      &    0.5113(5)      &    0.5121(6)      \\
$y$ (O3)               &    0.7851(4)      &    0.7863(4)      &    0.7873(4)      \\
$x$ (O4)               &    0.4290(6)      &    0.4271(6)      &    0.4276(6)      \\
$y$ (O4)               &    0.4221(4)      &    0.4216(4)      &    0.4234(4)      \\
\hline
$B$ (Sr)  ({\AA}$^2$)  &    1.12(2)        &    1.23(2)        &    1.37(2)        \\
$B$ (Er1) ({\AA}$^2$)  &    0.58(1)        &    0.67(1)        &    0.72(1)        \\
$B$ (Er2) ({\AA}$^2$)  &    0.61(1)        &    0.70(1)        &    0.76(1)        \\
$B$ (O)   ({\AA}$^2$)  &    0.35(5)        &    0.56(5)        &    0.77(6)        \\
\hline
$R_{\texttt{wp}}$               &    10.2           &    10.2           &    10.5  \\
$R_{\texttt{p}}$                &    7.43           &    7.51           &    7.70  \\
$\chi ^2$              &    1.73           &    1.77           &    1.48           \\
\hline
\end{tabular*}
\end{table}

The measured ZFC and FC effective PM moments are 9.13(1) $\mu_\texttt{B}$ and 9.14(1) $\mu_\texttt{B}$ per Er$^{3+}$ ion, respectively. Both values are almost the same within errors, but they are indeed smaller than the expected theoretical value $M^{\rm eff}_{\rm theo} \sim 9.58$ $\mu_{\rm B}$ of the ground state $^4I_{15/2}$ determined by the Hund's rules. This decrease is in agreement with the previous study \cite{Karunadasa2005} of polycrystalline SrEr$_2$O$_4$ and in addition may indicate that $\sim$4.6\% Er$^{3+}$ moments are frozen even in the high-temperature range of 200-300 K, or a small fraction of AFM spin interactions still exist in that temperature regime. Magnetization measurements at even higher temperatures would be of interest.

The deduced ZFC CW temperature $\theta^{\texttt{ZFC}}_{\texttt{CW}}$ = -15.94(3) K, which indicates a net AFM coupling strength and is by $\sim$2.4 K smaller than the corresponding value extracted from the polycrystalline SrEr$_2$O$_4$ \cite{Karunadasa2005}. This decrease in $\theta_{\rm CW}$ is $\sim$18.1\%, which is mainly due to the fact that the single-crystalline sample is more stoichiometric than the corresponding polycrystalline one because a single-crystalline sample naturally stays in the stablest state \cite{Li2008, Li2007-1, Li2007-2, Li2009}. It is noteworthy that the effect of FC the sample at 500 Oe renders a decrease in the CW temperature to $\theta^{\texttt{FC}}_{\texttt{CW}}$ = -16.69(3) K, by $\sim$4.71\%. This reduction results from two main contributions: one is $\sim$0.21\% decrease in the slope of $\chi^{-1}$, \emph{i.e.}, $\partial \chi^{-1} / \partial T$; another is $\sim$4.24\% increase in the intercept of $\chi^{-1}$ (at \emph{T} = 0 K). This interesting observation rules out the above hypothesis of the existence of frozen Er$^{3+}$ moments between 200 and 300 K and further supports that there indeed exist AFM couplings in that temperature regime. In addition, small applied magnetic fields (like 500 Oe used in this study) can stabilize such kind of weak AFM couplings before the processes of spin-flop and spin-flip transitions \cite{Quintero2012, Tian2010, Toft2012, Li2014SFP} occurring at higher field strengthes.

\subsection{Magnetization \emph{versus} applied magnetic field}

We measured the magnetization as a function of applied magnetic field with a powdered single-crystalline sample as shown in Fig.~\ref{Fig3}. At 2 K and 9 T, the measured magnetization $M^{\texttt{FC}}_{\texttt{mea}}$ = 4.3(5) $\mu_\texttt{B}$ (Table 1), roughly consistent with the previous study \cite{Karunadasa2005} of a powder sample and $\sim$47.8\% of the theoretical saturation value, 9.0 $\mu_\texttt{B}$, indicating that nearly half Er$^{3+}$ spin moments are magnetically frustrated. This is consistent with the nonlinear increase in the magnetization curve with applied magnetic field at 2 K as shown in Fig.~\ref{Fig3}. Such kind of modification in $\partial M / \partial H$ with field has a clear temperature dependence, and it still exists tenderly at 50 K.

\subsection{High-resolution SXRPD study}

\begin{figure}[h]
\centering \includegraphics[width = 0.44\textwidth] {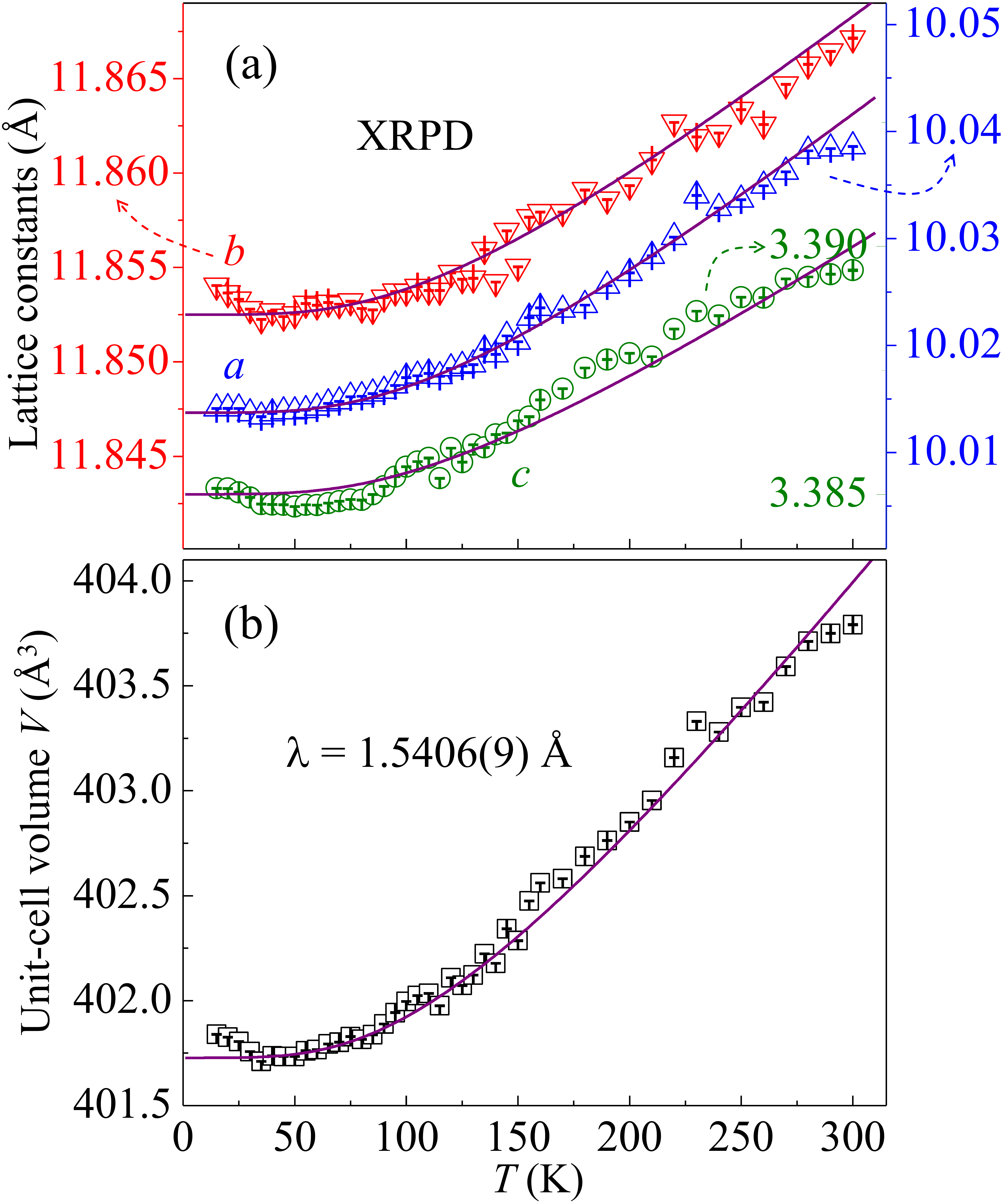}
\caption{
(a) Temperature variation of the lattice-constants, \emph{a}, \emph{b}, and \emph{c}. (b) The unit-cell volume, \emph{V}, expansion with temperature in the \emph{Pnam} symmetry. In this XRPD study, the temperature range is from 15 to 300 K. The data (symbols) are extracted by refining the collected XRPD patterns. Error bars are standard deviations obtained from the corresponding refinements. The solid lines are theoretical calculations of the temperature-dependent structural parameters using the Gr$\ddot{\texttt{u}}$neisen model with Debye temperature of $\theta_\texttt{D}$ = 460(10) K as described in the text.
}
\label{Fig5}
\end{figure}

To explore possible structural symmetry breaking in the compound SrEr$_2$O$_4$, we collected the high-resolution SXRPD patterns at 80 K, 298 K, and 500 K. Two representative patterns as well as the related structural refinements are displayed in Fig.~\ref{Fig4}. The refined results are listed in Table 2. The resulting crystal structure in one unit cell and the bent honeycombs at 80 K are schematically depicted in Figs.~\ref{Fig1}(a) and (b), respectively. All the observed Bragg peaks can be well indexed with the space group $Pnam$, and no extra peaks were detected. Therefore, there is no any structural phase transition between 80 and 500 K for the compound SrEr$_2$O$_4$. Indeed, the previous powder neutron-diffraction study \cite{Karunadasa2005} shows that the room-temperature structure of polycrystalline SrEr$_2$O$_4$ is well consistent with the space group $Pnam$.

The resulting lattice constants, \emph{a}, \emph{b}, and \emph{c} as listed in Table 2, almost increase linearly from 80 to 500 K. So does the corresponding unit-cell volume. This indicates that the change in lattice constants of SrEr$_2$O$_4$ is dominated mainly by the phonons' variation with temperature, and there is almost no electronic contribution. This is in agreement with our resistivity measurements, where the estimated resistance of the studied single crystal is larger than 10$^6$ ohm. Therefore, the compound SrEr$_2$O$_4$ is a robust insulator. The differences between refined room-temperature structural parameters of SrEr$_2$O$_4$ compound from our single-crystal study (as listed in Table 2) and from the previous study \cite{Karunadasa2005} of a polycrystalline sample (as shown in the Table 1 of Ref. 14) are probably due to different stoichiometries of the two types of sample state (polycrystal and single crystal) as foregoing remarks.

In principle, X-ray scattering is specifically sensitive to the distribution of outer electron clouds. As listed in Table 2, the refined Debye-Waller factors (\emph{B}) of the Sr, Er1, Er2, and especially the O ions, display an almost linear temperature dependence, which reflects the behavior of equilibrium atomic vibrations. This observation, to some extent, indicates that the valence electrons in SrEr$_2$O$_4$ are localized consistent with our resistivity measurements and the foregoing remarks.

\subsection{Thermal expansion studied by the temperature-dependent XRPD}

To explore the thermal expansions along the crystallographic \emph{a}, \emph{b}, and \emph{c} directions, we monitored a detailed temperature dependence of the XRPD pattern. The resulting temperature-dependent lattice constants (\emph{a}, \emph{b}, and \emph{c}) and unit-cell volume (\emph{V}) between 15 and 300 K are shown in Fig.~\ref{Fig5} (symbols).

Since SrEr$_2$O$_4$ is an insulator, we can neglect the electronic contribution (which is $\propto T^2$, but actually much smaller than the contribution from lattice vibrations) to the thermal expansion of the lattice configuration ($\varepsilon$). The temperature variation of the nonmagnetic contribution component is then mainly from phonons, which can approximately be estimated based on the Gr$\ddot{\texttt{u}}$neisen rules at zero pressure with the first-order fashion \cite{Wallace1998, Vocadlo2002, Brand2009, Li2012, Gavarri1988}:
\setlength\arraycolsep{1.4pt} 
\begin{eqnarray}
\varepsilon(T) = \varepsilon_0 + K_0U,
\label{Gr1}
\end{eqnarray}
where $\varepsilon_0$ is the lattice configuration at zero Kelvin, $K_0$ is a constant reflecting the compressibility of the sample, and the internal energy \emph{U} can be calculated based on the Debye approximations:
\setlength\arraycolsep{1.4pt} 
\begin{eqnarray}
U(T) = 9Nk_\texttt{B}T\mathrm{(}T/\Theta_\texttt{D}\mathrm{)}^3 \int^{\Theta_\texttt{D}/T}_0 \frac{x^3}{e^x - 1}dx,
\label{Gr2}
\end{eqnarray}
where \emph{N} (= 7) is the number of atoms per formula unit, $\Theta_\texttt{D}$ is the Debye temperature. With model eqn~(\ref{Gr1}) and (\ref{Gr2}), we fit the lattice configuration (\emph{a}, \emph{b}, \emph{c}, and \emph{V}) of SrEr$_2$O$_4$ in the temperature range from 150 to 300 K and extrapolated the fits down to low temperatures (15-150 K) as shown in Fig.~\ref{Fig5} (solid lines). As a whole, the temperature-dependent lattice constants comply well with the theoretical estimations. The fit to the unit-cell volume \emph{V} results in $\Theta_\texttt{D} = 460(10)$ K, $V_0 = 401.7(1)$ {\AA$^3$}, and $K = 4.8(5) \times 10^{19}$ {\AA$^3$} J$^{-1}$. It is pointed out that the upturn of the \emph{V}-curve below 25 K as shown in Fig.~\ref{Fig5}(b), which results largely from the corresponding increase in lattice constant \emph{b} (Fig.~\ref{Fig5}(a)), is attributed mainly to our X-ray powder diffractometer because this behavior normally indicates a formation of itinerant moments in the valence bands \cite{Li2012}, whereas such case is completely inconsistent with the fact that the compound SrEr$_2$O$_4$ is a robust insulator as foregoing remarks. We compare the lattice variations between 50 K and 280 K, \emph{i.e.}, $(a_{\texttt{280K}} - a_{\texttt{50K}})/a_{\texttt{50K}}$ = 0.243(1)\%, $(b_{\texttt{280K}} - b_{\texttt{50K}})/b_{\texttt{50K}}$ = 0.112(1)\%, and $(c_{\texttt{280K}} - c_{\texttt{50K}})/c_{\texttt{50K}}$ = 0.137(1)\%, which jointly results in $(V_{\texttt{280K}} - V_{\texttt{50K}})/V_{\texttt{50K}}$ = 0.492(1)\%. Therefore, the thermal expansion is anisotropic along the three crystallographic orientations (\emph{a}, \emph{b}, and \emph{c} axes).

\section{Conclusions}

In summary, we have studied a single crystal of Sr$_{1.04(3)}$Er$_{2.09(6)}$O$_{4.00(1)}$ by magnetization measurements and X-ray powder diffraction studies. By modeling the ZFC and FC magnetization with a CW law, we find that the CW temperature is negative, \emph{e.g.}, $\theta^{\texttt{ZFC}}_{\texttt{CW}}$ = -15.94(3) K, indicating a net AFM coupling strength. The difference between the refined ZFC and FC CW temperatures and the reduction of the relevant effective PM moments in contrast to the theoretical saturation value indicate that a small fraction of AFM couplings may persist up to 300 K. The measured magnetic moment per Er$^{3+}$ ion at 2 K and 9 T is 4.3(5) $\mu_\texttt{B}$, nearly half the corresponding theoretical saturation value (9 $\mu_\texttt{B}$), which implies that there exists a strong AFM frustration, especially at temperature points below $\sim$15 K, and that the measured moment may be from only one of the two inequivalent Er sites in accord with the previously-determined magnetic structure. Our high-resolution powder synchrotron X-ray diffraction studies demonstrate that the structural symmetry remains with the orthorhombic one in the studied temperature range from 80 to 500 K, and there is no any structural phase transition detected. The deduced temperature-dependent lattice parameters by our in-house XRPD studies display an anisotropic thermal expansion along the \emph{a}, \emph{b}, and \emph{c} axes and agree well with the Gr$\ddot{\texttt{u}}$neisen rules consistent with the localized magnetic configuration.

\section{Acknowledgements}

H.F.L is grateful to instrument scientists at the I11 beamline located at Diamond Light Source Ltd., United Kingdom, for expert technical assistance.


\end{document}